\documentclass[prb,aps,twocolumn,amsmath,amssymb,floatfix,showpacs,
superscriptaddress]{revtex4}

\usepackage[dvips]{graphics}
\usepackage{color}
\definecolor{gold}{rgb}{0.85,0.66,0}
\definecolor{dred}{rgb}{0.6,0,0}
\definecolor{dblue}{rgb}{0,0,0.8}

\begin{document}

\title{\textcolor{dblue}{A proposal for the measurement of Rashba and 
Dresselhaus spin-orbit interaction strengths in a single sample}}

\author{Santanu K. Maiti}

\email{santanu@post.tau.ac.il}

\affiliation{School of Chemistry, Tel Aviv University, Ramat-Aviv,
Tel Aviv-69978, Israel}

\author{Shreekantha Sil}

\affiliation{Department of Physics, Visva-Bharati, Santiniketan, West
Bengal-731 235, India}

\author{Arunava Chakrabarti}

\affiliation{Department of Physics, University of Kalyani, Kalyani,
West Bengal-741 235, India}

\begin{abstract}

We establish an exact analytical treatment for the determination of
the strengths of the Rashba and Dresselhaus spin-orbit interactions in
a single sample by measuring persistent spin current. A hidden symmetry
is exploited in the Hamiltonian to show that the spin current vanishes
when the strength of the Dresselhaus interaction becomes equal to the
strength of the Rashba term. The results are sustained even in the
presence of disorder and thus an experiment in this regard will be
challenging.

\end{abstract}


\maketitle

Spintronics is a rapidly growing field that aims at the achievement of  
efficient quantum devices, such as the magnetic memory circuits and 
computers, in which one needs to manipulate the spin of the electron  
rather than its charge~\cite{zutic,datta,ando,ding,bellucci,aharony1,
aharony2}. Magnetic nano-structures and quasi-one dimensional semiconductor 
rings have been acknowledged as ideal candidates for testing the effects 
of quantum coherence in low dimensions, and have been extensively 
investigated to test their potential as new generation quantum devices 
mentioned above~\cite{moldo,engels}.

Spin-orbit interaction (SOI) in semiconductors is a central mechanism 
that determines the essential physics in the meso- and nano-scales, and 
is largely responsible for the prospect of semiconducting structures as 
potential quantum devices. Rashba spin-orbit interaction (RSOI) and the 
Dresselhaus spin-orbit interaction (DSOI) are the two typical spin-orbit 
interactions that one encounters in a conventional semiconductor. 

The Rashba spin-orbit field in a solid is attributed to an electric field 
that originates from a structural inversion asymmetry whereas, the 
Dresselhaus interaction comes from bulk inversion asymmetry~\cite{meier}.  
Quantum rings formed at the interface of two semiconducting materials 
are ideal candidates to unravel the interplay of the two kinds of SOI. A 
quantum ring in a heterojunction, formed by trapping a two-dimensional gas 
of electrons in a quantum well, generates a  {\it band offset} at the 
interface of two different semiconducting materials. This creates an 
electric field. This electric field is described by a potential gradient 
normal to the interface~\cite{premper}. The potential at the interface is 
asymmetric, leading to the presence of a RSOI. On the other hand, at such 
interfaces, the bulk inversion symmetry is naturally broken, and DSOI plays 
its part.

Needless to say, an accurate estimation of the SOI's is crucial in the 
field of spintronics. The RSOI can be controlled by a gate voltage 
placed in the vicinity of the sample~\cite{engels,meier,premper,cmhu,
grundler}. Thus, in principle, all possible values of the RSOI can be 
achieved. Measurement of RSOI has already been reported in the 
literature~\cite{nitta}. Comparatively speaking, reports on the techniques 
of measurement of the DSOI are relatively few, and are 
mainly based on an optical monitoring of the spin precession of the 
electrons~\cite{meier,studer}, measurement of electrical conductance 
of nano-wires~\cite{scheid,jaya}, or photo-galvanic methods~\cite{yin}.

In the present communication we propose a method to measure the strengths 
of both the RSOI and DSOI by measuring the spin current flowing through a 
single sample. An important issue, while the measurement of a spin current 
\begin{figure}[ht]
{\centering \resizebox*{2.5cm}{2.5cm}{\includegraphics{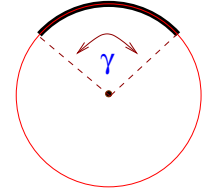}}\par}
\caption{(Color online). Schematic view of a mesoscopic ring where a 
section of the ring (shaded region measured by the angle $\gamma$) is 
subjected to SOI, while the other section is free from the SOI.}
\label{ring}
\end{figure}
in a mesoscopic system is concerned, is the non-conservation of spin 
caused by the ubiquitous presence of the spin-orbit coupling. A proper 
definition of spin current is therefore urgent~\cite{wang} and, its 
measurement in a sample is challenging. It  has only recently been possible, 
using a novel version of the Doppler effect, to quantify the spin current 
in a ferromagnetic wire~\cite{vlaminck}, and by using spin relaxation 
modulation induced by spin injection~\cite{ando1,ando2}. 

Inspired by the success in the measurement of spin current we provide a 
completely analytical scheme that immediately  suggests an experiment 
to measure the strengths of either RSOI or the DSOI in the same sample. 
Our system is a mesoscopic ring grafted at a heterojunction, but in such 
a manner that the spin-orbit interaction is effective only in a fraction 
of the perimeter of the ring (see Fig.~\ref{ring}). The spin current 
should be measured in the SO interaction-free region where the spin flip 
scattering is absent. One thus need not worry about the non-conservation 
of spin and the spin current in any `bond' in the non-shaded region in 
Fig.~\ref{ring} is the same, and becomes a well defined quantity.

In this letter we prove that, in such a mesoscopic ring the spin 
current will be zero whenever the strengths of the RSOI and the DSOI are 
equal. As we have already mentioned above, the RSOI strength can in 
principle, be smoothly varied with applied gate voltage. This leads to 
the idea of estimating the strength of the DSOI if we happen to know 
a measured value of the RSOI. The idea is as follows.
We devise a unitary transformation that acts on the full Hamiltonian 
of the system to extract a subtle symmetry that is finally exploited to 
achieve the result. Incidentally, similar kind of symmetry has previously 
been reported in literature~\cite{schliemann,andrei,brataas}. In this work 
we have focused on the effect of the symmetry on the spin current for a 
system in the presence of both the Rashba and the Dresselhaus interactions.
Our result is analytically exact, and is true for any 
value of the spin-orbit interactions. It should therefore be observable in 
a suitably designed experiment. In addition, we numerically calculate the 
persistent spin current in the system, and determine its dependence on the 
length of the sample involving the SOI (determined by the angle 
$\gamma$ in Fig.~\ref{ring}). The numerical results corroborate our 
analytical work, and remain robust even in the presence of disorder.

\vskip 0.1cm
\noindent
$\bullet$ {\bf {Analytical treatment for the determination of 
SOI strengths:}} Let us refer to Fig.~\ref{ring}. Only the bold portion 
of the arc contains the spin-orbit interactions. Within a tight-binding 
framework the Hamiltonian for such a ring reads,
\begin{equation}
\mbox{\boldmath $H$} = \mbox{\boldmath $H_0$} -i\alpha \mbox{\boldmath 
$H_{RSO}$} + i\beta \mbox{\boldmath $H_{DSO}$}
\end{equation}
where,
\begin{equation}
\mbox{\boldmath $H$}_0 = \sum_n \mbox{\boldmath $c_n^{\dagger} 
\epsilon_n c_n$} + \sum_n \left(\mbox{\boldmath $c_{n+1}^{\dagger} t$} 
\mbox {\boldmath $c_n$} + h.c. \right).
\label{ham0}
\end{equation}
The Rashba and the Dresselhaus spin-orbit parts of the Hamiltonian, viz, 
\mbox{\boldmath $H_{RSO}$} and \mbox{\boldmath $H_{DSO}$}, are given by, 
\begin{eqnarray}
\mbox{\boldmath $H_{RSO}$}  & = &   
\sum_n \left(\mbox{\boldmath $c_{n}^{\dagger}$}
\mbox{\boldmath$\sigma_x$} \cos\varphi_{n,n+1}  
\mbox{\boldmath $c_{n+1}$} \right. \nonumber \\
& + & \left. \mbox{\boldmath $c_{n}^{\dagger}$}
\mbox{\boldmath$\sigma_y$} \sin\varphi_{n,n+1} 
\mbox{\boldmath $c_{n+1}$} + h.c. \right) \nonumber \\
\mbox{\boldmath $H_{DSO}$}  & = &   
\sum_n \left(\mbox{\boldmath $c_{n}^{\dagger}$}
\mbox{\boldmath$\sigma_y$} \cos\varphi_{n,n+1}  
\mbox{\boldmath $c_{n+1}$} \right. \nonumber \\
& + & \left. 
\mbox{\boldmath $c_{n}^{\dagger}$}
\mbox{\boldmath$\sigma_x$} \sin\varphi_{n,n+1} 
\mbox{\boldmath $c_{n+1}$} + h.c. \right) 
\label{equ1}
\end{eqnarray}
where, $n=1$, $2$, $\dots$, $N$ is the site index along the azimuthal
direction $\varphi$ of the ring. The other factors in Eq.~\ref{equ1}
are as follows.\\
\mbox{\boldmath $c_n$}=$\left(\begin{array}{c}
c_{n \uparrow} \\
c_{n \downarrow}\end{array}\right);$
\mbox{\boldmath $\epsilon_n$}=$\left(\begin{array}{cc}
\epsilon_{n\uparrow} & 0 \\
0 & \epsilon_{n\downarrow} \end{array}\right);$ 
\mbox{\boldmath $t$}=$t\left(\begin{array}{cc}
1 & 0 \\
0 & 1 \end{array}\right)$. \\
~\\
\noindent
Here $\epsilon_{n\sigma}$ is the site energy of an electron at the $n$-th 
site of the ring with spin $\sigma$ ($\uparrow,\downarrow$). $t$ is the 
nearest-neighbor hopping integral. $\alpha$ and $\beta$ are the isotropic 
nearest-neighbor transfer integrals which measure the strengths of Rashba 
and Dresselhaus SOI, respectively, and 
$\varphi_{n,n+1}=\left(\varphi_n+\varphi_{n+1}\right)/2$, where
$\varphi_n=2\pi(n-1)/N$. \mbox{\boldmath $\sigma_x$},
\mbox{\boldmath $\sigma_y$} and \mbox{\boldmath $\sigma_z$} are the Pauli
spin matrices. $c_{n \sigma}^{\dagger}$ ($c_{n \sigma}$) is the creation
(annihilation) operator of an electron at the site $n$ with spin $\sigma$
($\uparrow,\downarrow$). The form of the Hamiltonian written in the case 
of a ring is discussed elsewhere~\cite{santanu}.

We now define a transformation, 
\begin{equation}
\mbox{\boldmath $U$} = \mbox{\boldmath $\sigma_{z} \left(\frac{\sigma_{x} 
+ \sigma_{y}}{\sqrt{2}} \right)$} 
\label{transform}
\end{equation}
which, by definition is unitary. It is simple to verify that, 
\begin{equation}
\mbox{\boldmath $U\sigma_x U^{\dagger}$}= -\mbox{\boldmath $\sigma_y$};\, 
\mbox{\boldmath $U\sigma_y U^{\dagger}$}= -\mbox{\boldmath $\sigma_x$};\,
\mbox{\boldmath $U\sigma_z U^{\dagger}$}= -\mbox{\boldmath $\sigma_z$}.
\label{newsigma}
\end{equation}
Using the operator defined in Eq.~\ref{transform} we now make a change of 
basis and describe the full Hamiltonian in terms of the operators 
$\mbox{\boldmath $\tilde c_n$} = \mbox{\boldmath $U c_n$}$, and 
$\mbox{\boldmath $\tilde c_n^{\dagger}$} = 
\mbox{\boldmath $c_n^{\dagger}U^{\dagger}$}$.
The Hamiltonian in the new basis reads, 
\begin{equation}
\mbox{\boldmath $\tilde H$} = \mbox{\boldmath $\tilde H_0$} - i\beta 
\mbox{\boldmath $\tilde H_{RSO}$} + i\alpha \mbox{\boldmath 
$\tilde H_{DSO}$}
\label{newham}
\end{equation}
where,
\begin{equation}
\mbox{\boldmath $\tilde H_0$}=\sum_n \mbox{\boldmath $\tilde c_n^{\dagger} 
\epsilon_n \tilde c_n$} + \sum_n \left(\mbox{\boldmath $\tilde 
c_{n+1}^{\dagger} t$} \mbox {\boldmath $\tilde c_n$} + h.c. \right)
\label{newham0}
\end{equation}
and, the spin-orbit parts of the Hamiltonian in this new basis are 
given by, 
\begin{eqnarray}
\mbox{\boldmath $\tilde H_{RSO}$}  & = &   
\sum_n \left(\mbox{\boldmath $\tilde c_{n}^{\dagger}$}
\mbox{\boldmath$\sigma_x$} \cos\varphi_{n,n+1}  
\mbox{\boldmath $\tilde c_{n+1}$} \right. \nonumber \\
& + & \left. \mbox{\boldmath $\tilde c_{n}^{\dagger}$}
\mbox{\boldmath$\sigma_y$} \sin\varphi_{n,n+1} 
\mbox{\boldmath $\tilde c_{n+1}$} + h.c. \right) \nonumber \\
\mbox{\boldmath $\tilde H_{DSO}$}  & = &   
\sum_n \left(\mbox{\boldmath $\tilde c_{n}^{\dagger}$}
\mbox{\boldmath$\sigma_y$} \cos\varphi_{n,n+1}  
\mbox{\boldmath $\tilde c_{n+1}$} \right. \nonumber \\
& + & \left. \mbox{\boldmath $\tilde c_{n}^{\dagger}$}
\mbox{\boldmath$\sigma_x$} \sin\varphi_{n,n+1} 
\mbox{\boldmath $\tilde c_{n+1}$} + h.c. \right).
\label{newequ1}
\end{eqnarray}
A simultaneous look at the Eqs.~\ref{newham} and \ref{newequ1} reveal that
in the new basis, the strengths of the Rashba and Dresselhaus spin-orbit 
interactions viz, $\alpha$ and $\beta$ have been interchanged. 
As a consequence, the polarized spin current operator in the quantized 
direction ($+Z$), for the free region, defined as 
$\mbox{\boldmath $J_{s}^z$}=\frac{1}{2}\left(\mbox{\boldmath ${\sigma_z}$}
\mbox{\boldmath ${\dot{x}}$}+ \mbox{\boldmath ${\dot{x}}$} \mbox{\boldmath 
${\sigma_z}$}\right)$ flips its sense of circulation, and becomes 
$-\mbox{\boldmath $J_{s}^z$}$. Here, \mbox{\boldmath $x$}$=\sum n 
\mbox{\boldmath$c_n^{\dagger} c_n$}$. This immediately leads to the 
most important observation that, as soon as $\alpha = \beta$, 
\mbox{\boldmath$H$} and \mbox{\boldmath $\tilde H$} become identical. 
That is, the full Hamiltonian remains invariant under the unitary 
transformation defined by Eq.~\ref{transform} whenever the strengths 
of the Rashba and the Dresselhaus spin-orbit interactions become equal. 
The energy eigenvalues obviously remain unchanged in this case. 

At the same time, from the last equation in Eq.~\ref{newsigma} we see that 
$\mbox{\boldmath $U\sigma_z U^{\dagger}$} = -\mbox{\boldmath $\sigma_z$}$ 
which means that, the spin current $J_s^z$ calculated in the original basis 
{\it becomes exactly equal} to the spin current $-J_s^z$ calculated in the 
new basis. This can happen only when $J_s^z = 0$ as $\alpha = \beta$. The 
RSOI can be 
controlled by a gate voltage, and hence its strength is determined. So, an 
experiment on the measurement of spin current in a mesoscopic ring, where 
the RSOI is continuously varied, will show a complete disappearance of the 
spin current as soon as the strength of the DSOI becomes equal to that of 
the RSOI. We thus have a means of determining DSOI from a measurement of 
the spin current. The spin current, however, as stated earlier, is to be 
measured in that region of the ring which is free from the spin-orbit 
coupling.

Needless to say, the strength of the Rashba interaction can be obtained 
by the same method, if we know the Dresselhaus interaction beforehand.

\vskip 0.1cm
\noindent
$\bullet$ {\bf {Calculation of persistent spin current:}}
In the second quantized form the spin current operator 
\mbox{\boldmath $J_s^z$} for the free region can be written as,
\begin{eqnarray}
\mbox{\boldmath $J_{s}^z$} &=& 2 i\pi t\sum_n \left(
\mbox{\boldmath $c_n^{\dagger}\sigma_z$} \mbox{\boldmath $c_{n+1}$} - 
\mbox{\boldmath $c_{n+1}^{\dagger}\sigma_z$} \mbox{\boldmath $c_n$} \right).
\label{equ9}
\end{eqnarray}
Therefore, for a particular eigenstate $|\psi_k\rangle$ the persistent
spin current becomes,
$J_s^{z,k}=\langle \psi_k|\mbox{\boldmath$J_s^z$}|\psi_k\rangle$,
where $|\psi_k\rangle=\sum\limits_p a_{p,\uparrow}^k |p\uparrow\rangle
+ a_{p,\downarrow}^k |p\downarrow\rangle$. Here $|p\uparrow\rangle$'s
and $|p\downarrow\rangle$'s are the Wannier states and $a_{p,\uparrow}^k$'s
\begin{figure}[ht]
{\centering \resizebox*{8cm}{7cm}{\includegraphics{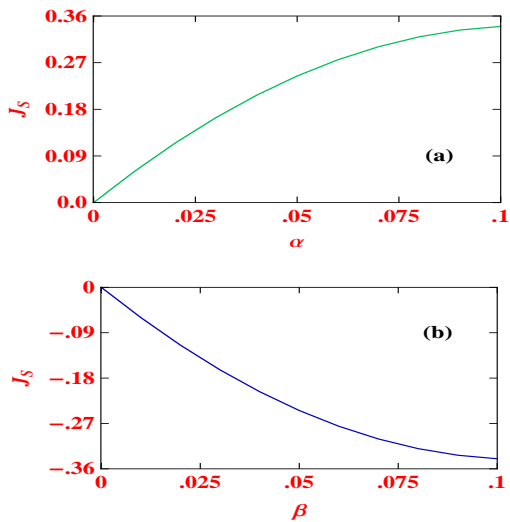}}\par}
\caption{(Color online). Persistent spin current as a function of SO 
coupling strength for an ordered $40$-site half-filled ring, when 
$\gamma$ is set at $\pi$. (a) $\beta=0$ and (b) $\alpha=0$.}
\label{current1}
\end{figure}
and $a_{p,\downarrow}^k$'s are the corresponding coefficients. In terms of
these coefficients, the final expression of persistent spin current for 
$k$-th eigenstate reads,
\begin{eqnarray}
J_s^{z,k} &=& 2\pi i t \sum_n\left\{a_{n,\uparrow}^{k\,*}\,a_{n+1,\uparrow}^k
- a_{n+1,\uparrow}^{k\,*}\, a_{n,\uparrow}^k \right\}
\nonumber \\
& - & 2\pi i t \sum_n \left\{a_{n,\downarrow}^{k\,*}\, a_{n+1,\downarrow}^k - 
a_{n+1,\downarrow}^{k\,*}\, a_{n,\downarrow}^k \right\}.
\label{equ44}
\end{eqnarray}
Let us rename the polarized spin current $J_s^{z,k}$ as $J_s^k$ for the 
sake of simplicity. At absolute zero temperature ($T=0$\,k), net persistent 
spin current in a mesoscopic ring for a particular filling 
can be obtained by taking the sum of individual contributions from the 
energy levels with energies less than or equal to Fermi energy $E_F$. 
Therefore, for $N_e$ electron system total spin current becomes
$J_s=\sum_k^{N_e} J_s^k.$

We measure spin current in the region which is free from the SOI, 
and, since the spin currents between any two neighboring sites in this 
interacting free region are identical to each other we compute the current 
only in a single bond.

In the present work we compute all the essential features of persistent
spin current and related issues at absolute zero temperature and choose 
the units where $c=h=e=1$. Throughout our numerical work we fix $t=1$ 
and measure the energy scale in unit of $t$.

\vskip 0.1cm
\noindent
$\bullet$ {\bf {Numerical results:}}
In Fig.~\ref{current1}, we present the variation of the spin current as 
a function of the RSOI and the DSOI for a $40$-site ordered half-filled 
ring. The angle $\gamma = \pi$. Figs.~\ref{current1}(a) and (b) refer to 
\begin{figure}[ht]
{\centering \resizebox*{7.25cm}{4.25cm}{\includegraphics{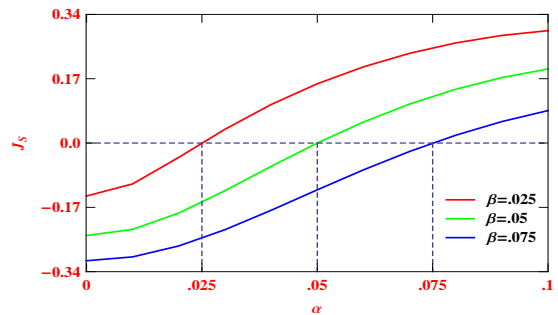}}\par}
\caption{(Color online). Persistent spin current as a function of Rashba
SO interaction strength for a $40$-site ordered ring in the half-filled 
case considering different values of $\beta$, when $\gamma$ is fixed at 
$\pi$.}
\label{current2}
\end{figure}
\begin{figure}[ht]
{\centering \resizebox*{7.25cm}{4.25cm}{\includegraphics{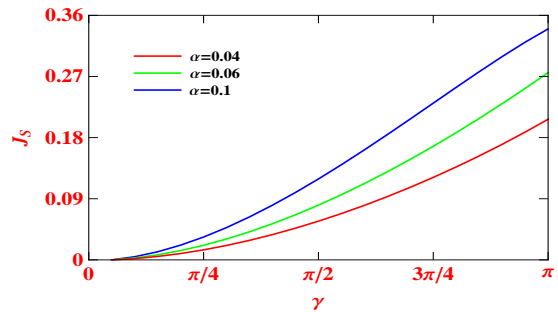}}\par}
\caption{(Color online). Persistent spin current as a function of $\gamma$ 
for a $40$-site ordered ring in the half-filled case considering different 
values of $\alpha$, when $\beta$ is fixed at $0$.}
\label{current3}
\end{figure}
the cases where only the RSOI and the DSOI are present respectively. The 
RSOI and the DSOI drive the spin current in opposite directions (with 
equal magnitudes when $\alpha = \beta$), as is evident from the figure.

Figure~\ref{current2} shows the variation of the spin current as a function
of the strength of the RSOI, plotted for various values of the strength of 
the DSOI. This figure is a clear demonstration of the fact that, whenever 
the strength of the DSOI becomes equal to that of the RSOI, the spin 
current becomes zero. We again emphasize that, the vanishing of the 
persistent spin current will be observed for any strength of the RSOI, 
provided it equals the DSOI strength. Hence, it is expected to be 
observed in experiments as well. The region of 
the mesoscopic ring in which the SOI is `on', plays an important role in 
determining the strength of the spin current in the system. To get an idea, 
we systematically study the variation of $J_s$ as a function of $\gamma$ 
with the DSOI $\beta = 0$. For a given value of the angle $\gamma$, the 
current increases with increasing values of the RSOI strength $\alpha$ 
(see Fig.~\ref{current3}).

Finally, in view of a possible experiment, we test the robustness of our 
results by considering a $40$-site disordered ring. Disorder is introduced
via a random distribution (width $W=1$) of the values of the on-site 
\begin{figure}[ht]
{\centering \resizebox*{7.25cm}{4.25cm}{\includegraphics{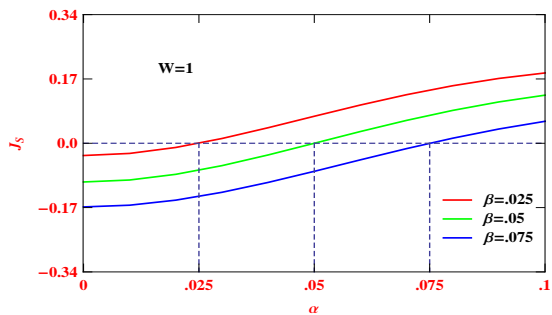}}\par}
\caption{(Color online). Persistent spin current as a function of Rashba
SO interaction strength for a $40$-site disordered ($W=1$) ring in the 
half-filled case considering different values of $\beta$, when $\gamma$ 
is fixed at $\pi$.}
\label{current5}
\end{figure}
potentials (diagonal disorder), and results averaged over $40$ disorder
configurations have been presented (see Fig.~\ref{current5}). Since 
\mbox{\boldmath $H_0$} remains invariant under the unitary transformation 
defined by Eq.~\ref{transform}, all the qualitative results should remain 
unchanged even in the presence of disorder. However, the disorder will 
reduce the amplitude of the spin current. This is precisely what we find 
in our numerical analysis. The current still becomes zero as soon as 
$\alpha=\beta$. This observation strengthens our claim that a suitably 
designed experiment on the measurement of the spin current will lead to 
an exciting method of measurement of the Dresselhaus spin-orbit interaction 
from a knowledge of the gate controlled Rashba spin-orbit interaction in 
a mesoscopic ring.

A relevant question in view of the above discussion is 
the possibilities of the experimental detection of persistent spin current. 
The spin current, as obtained from the theoretical calculations already 
existing in the literature as well as from the present calculation, is a 
robust phenomenon undisturbed by the presence of disorder. Determination 
of spin current is also very much possible, and already a few experiments 
have been carried out in this direction~\cite{expt1,expt2,expt3,expt4}.
For example, by studying the Kerr effect~\cite{expt1,expt2,expt3} associated
with spin accumulations induced by the spin current or by investigating the
reciprocal spin Hall effect~\cite{expt4} persistent spin current can be 
detected. It is also well known that a spin current may produce a spin 
torque which can be measured experimentally~\cite{expt5,expt6,expt7}. This 
definitely provides a way of estimating the spin current. Probably a more 
convenient way of detecting persistent spin current is the measurement of 
the electric field and electric potential induced by it~\cite{expt8,expt9,
expt10,expt11,expt12}. This is quite analogous to the detection of 
persistent charge current in a mesoscopic ring by measuring its induced 
magnetic field~\cite{expt13,expt14}. We can also measure the strengths of 
spin-orbit interactions by attaching two electrodes in a mesoscopic ring. 
In that case also the persistent spin current as well as the transport spin 
current have well defined expressions and simply by measuring the transport 
spin current we can have an estimate of the SO interaction strengths. These 
latter observations are the major issues of our forthcoming paper.

In conclusion, we  present an exact analytical method which shows that 
the strengths of the Rashba or the Dresselhaus spin-orbit interactions 
can be determined in a single mesoscopic ring by noting the vanishing of 
the spin current in the sample. A unitary transformation is prescribed, 
which when applied to the spin-orbit Hamiltonian brings out a hidden 
symmetry. The symmetry is exploited to prove that, by making the strengths 
of the two interactions equal, one achieves a zero spin current in the 
system. We provide numerical results which support all our analytical 
findings, and show that the vanishing of the spin current is a robust 
effect even in the presence of disorder. This last observation gives 
us confidence to propose an experiment in this line.

\end{document}